\documentclass[twoside,fleqn]{article}
\usepackage{espcrc2}
\usepackage{graphicx,cite}

\newcommand{\figwidth}{8.0cm}
\newcommand{\figheight}{4.5cm}

\newcommand{\nn}{\nonumber}
\newcommand{\IM}{\mbox{\rm Im}}
\newcommand{\eqn}[1]{(\ref{#1})}

\newcommand{\gev}{\mbox{\rm GeV}}

\newcommand{\MSb}{{\overline{\rm MS}}}
\newcommand\lsim{\mathrel{\rlap{\lower4pt\hbox{\hskip1pt$\sim$}}
    \raise1pt\hbox{$<$}}}
\newcommand\gsim{\mathrel{\rlap{\lower4pt\hbox{\hskip1pt$\sim$}}
    \raise1pt\hbox{$>$}}}

\newcommand{\npb}[3]{{\it Nucl. Phys. }{\bf B #1} (#2) #3}

\newcommand{\plb}[3]{{\it Phys. Lett. }{\bf B #1} (#2) #3}

\newcommand{\prd}[3]{{\it Phys. Rev. }{\bf D #1} (#2) #3}
\newcommand{\prl}[3]{{\it Phys. Rev. Lett. }{\bf #1} (#2) #3}
\newcommand{\prep}[3]{{\it Phys. Rep. }{\bf #1} (#2) #3}

\newcommand{\hepph}[1]{{\tt hep-ph/#1}}

\title{
\vspace{-1.0cm}
{\sf \small \rightline{IFIC/02-39, FTUV/02-0903}}
\bigskip
{\Large \bf Charm and bottom quark masses from QCD moment sum rules}}
\author{M. Eidem\"uller\thanks{Talk given at the High-Energy Physics
	International Conference in Quantum Chromodynamics
        (QCD 2002), Montpellier, July 2002}
\address{
       {\em Departament de F\'{\i}sica Te\`orica, IFIC,
       Universitat de Val\`encia -- CSIC,}\\ 
       {\em Apt. Correus 22085, E-46071 Val\`encia, Spain}}
	}
\begin{document}

\begin{abstract}
\noindent
In this work the charm and bottom quark masses are determined from QCD moment
sum rules for the charmonium and upsilon systems. In our analysis
we include both the results from non-relativistic QCD and perturbation theory
at next-next-to-leading order. 
For the pole masses we obtain  $M_c=1.75\pm 0.15$~GeV and $M_b=4.98\pm 0.125$~GeV.
Using the potential-subtracted mass in intermediate steps of the
calculation the $\MSb$-masses are determined to $m_c(m_c) = 1.19 \pm 0.11$~GeV 
and $m_b(m_b) = 4.24 \pm 0.10$~GeV.
\end{abstract}

\maketitle



\section{Introduction}

An important task within modern particle phenomenology consists in the
determination of the quark masses, being fundamental parameters of the 
Standard Model.
In the past, QCD moment sum rule
analyses have been successfully applied for extracting the charm and
bottom quark masses from experimental data on the charmonium and
bottomium systems respectively \cite{svz:79,rry:85,n:89}.
The basic quantity in these investigations is the
vacuum polarisation function $\Pi(q^2)$:
\begin{eqnarray}
  \label{eq:1.a}
        \Pi_{\mu\nu}(q^2) &=& i \int d^4 x \ e^{iqx}\, \langle
        T\{j_\mu(x) j_\nu^\dagger(0)\}\rangle\nn\\
          &=& (q_\mu q_\nu-g_{\mu\nu}q^2)\,\Pi(q^2)\,,
\end{eqnarray}
where the relevant vector current is represented either by the charm 
$j_\mu^c(x)=(\bar{c}\gamma_\mu c)(x)$ or the bottom current
$j_\mu^b(x)=(\bar{b}\gamma_\mu b)(x)$.
Via the optical theorem, the experimental cross 
section $\sigma(e^+ e^- \to c\bar{c},b\bar{b})$
is related to the imaginary part of $\Pi(s)$:
\begin{displaymath}
  \label{eq:1.b}
	  R(s)=\frac{1}{Q_{c,b}^2}\,
	\frac{\sigma(e^+ e^- \to c\bar{c},b\bar{b})}
	  {\sigma(e^+ e^- \to \mu^+ \mu^-)}=12\pi\,
	\IM\, \Pi(s)\,.
\end{displaymath}
Usually, moments of the vacuum polarisation are defined by taking derivatives
of the correlator at $s=0$. However, in this work we allow for an
arbitrary evaluation point $s=-4 m^2 \xi$ to define the dimensionless 
moments \cite{e:02}:
\begin{eqnarray}
  \label{eq:1.c}
 {\cal M}_{n}(\xi) &=& \frac{12\pi^2}{n!} \left(4m^2 \frac{d}{ds}\right)^n
  \Pi(s)\bigg|_{s=-4 m^2 \xi} \nn\\	
	&=& 2 \int\limits_0^1 \!dv \, \frac{v(1-v^2)^{n-1}R(v)}
	{(1+\xi(1-v^2))^{n+1}} \,,
\end{eqnarray}
where $v=\sqrt{1-4m^2/s}$ is the velocity of the heavy quark. 
The parameter $\xi$ encodes much
information about the system. By taking  $\xi$ larger the evaluation
point moves further away from the threshold region. Consequently, the
theoretical expansions show a better convergence, but at the same time 
the sensitivity on the mass is reduced. 
The moments can either be calculated theoretically, including Coulomb
resummation, perturbation theory and nonperturbative contributions, or be
obtained from experiment. In this way one can relate the heavy quark masses
to the hadronic properties of the quark-antiquark systems.

A characteristic feature of these heavy-heavy bound state systems is
the Coulomb-like form of the potential. Developing the quantum mechanical
sum rules for Coulomb systems, in \cite{e:02} it has been shown that the 
application of fixed-order perturbation theory in such systems results in
unstable predictions for the masses. To obtain a stable sum rule it is
necessary to incorporate the threshold behaviour which can be calculated
in the framework of non-relativistic QCD (NRQCD). 

A natural choice for the mass in eq.~\eqn{eq:1.c} is the pole mass $M$. 
First we will employ the pole mass scheme to extract the
pole masses. However, as the pole masses suffer from renormalon ambiguities
\cite{b:99}, we shall then use the potential-subtracted (PS-) mass 
$m_{PS}$ \cite{b:98} to extract the ${\rm \MSb}$-masses.


\section{Coulomb resummation}

The theory of NRQCD provides a consistent framework
to treat the problem of heavy quark-antiquark production close to
threshold. 
The contributions can be described by a nonrelativistic Schr\"odinger
equation and systematically calculated in time-independent perturbation
theory (TIPT). 
The correlator is expressed in terms of a Green's function 
$G(k)=G(0,0,k)$ \cite{ps:91,h:98,pp:99}:
\begin{equation}
  \label{eq:3.a}
  \Pi(s)=\frac{N_c}{2M^2}\left(C_h(\alpha_s)G(k)+\frac{4k^2}{3M^2}G_C(k)\right)\,,
\end{equation}
where $k=\sqrt{M^2-s/4}$ and $M$ represents the pole mass.
The constant $C_h(\alpha_s)$ is a perturbative coefficient
needed for the matching between the full and the nonrelativistic 
theory and naturally depends on the hard scale.

To calculate the moments from the Green's function we will directly
perform the derivatives at $s=-4 M^2\xi$ according to eq. \eqn{eq:1.c}.
Since the Green's function is known analytically \cite{pp:99}
as a function of $k=k(s)$, this can be done numerically. In this way
we take advantage of the fact that the perturbative expansion parameter
depends on the evaluation point. 
In addition, we can extract the spectral density above threshold
by taking the imaginary part of the Green's function \cite{e:02,ej:01}.

The moments depend on the three scales $\mu_{soft}$, $\mu_{fac}$ and
$\mu_{hard}$, the soft, factorisation and hard scale respectively.
The residual dependence on these scales will turn out to give the
dominant error in the determination of the masses.
The large corrections are partly due to the definition of the pole mass.
These contributions can
be reduced by using an intermediate mass definition. In this analysis
we will use the potential-subtracted (PS) mass \cite{b:98} where
the potential below a separation scale $\mu_{sep}$ is subtracted.
This mass definition
leads to an improved scale dependence and a more precise
determination of the ${\rm \MSb}$-masses.


\section{Perturbative expansion}

The perturbative spectral function $R^{Pert}(s)$ can be expanded in 
powers of the strong coupling constant,
\begin{displaymath}
  \label{eq:4.a}
  R^{Pt}(s)= R^{(0)}(s)+ \frac{\alpha_s}{\pi}\,R^{(1)}(s)+
  \frac{\alpha_s^2}{\pi^2} \,R^{(2)}(s)+\ldots
\end{displaymath}
From this expression the corresponding moments
${\cal M}_n$ can be calculated via the integral of eq. \eqn{eq:1.c}.
The first two terms are known analytically and can for example be
found in ref. \cite{jp:97}.
$R^{(2)}(s)$ is still not fully known
analytically. We employ a method based on  
Pad{\'e}-approximants to construct the spectral density in the
full energy range \cite{cks:96,cks:97}. It uses available
information around $q^2=0$, at threshold and in the high
energy region. It has the advantage that it gives a good
description until relatively close to threshold, 
the region on which the quark
masses are most sensitive. 


\section{Condensate contributions}

The non-perturbative effects on the vacuum correlator are parametrised
by the condensates. The leading correction is the gluon condensate
contribution which is known up to next-to-leading order \cite{bbifts:94}.
Furthermore, in \cite{nr:83:1,nr:83:2} the dimension 6 and 8
condensates have been calculated. 
From the numerical analysis it turns out that the absolute contribution 
of the condensates to the moments is negligible for the bottomium and
of little influence for the charmonium. The  relative suppression of
the gluon condensate to former charmonium analyses is due to three
reasons: First, the absolute value of the
theoretical moments increases from the Coulomb resummation. 
Then we evaluate the moments at larger $\xi$ and smaller $n$ 
where the nonperturbative 
contributions are relatively small. Finally, since we obtain a larger 
pole mass than former analyses, the condensates,
starting with a power of $1/M^4$, are suppressed further.


\section{Phenomenological spectral function}

Experimentally, the six lowest lying $\psi-$ and $\Upsilon-$resonances 
have been observed. Furthermore, recent measurements of BES \cite{BES:01}
in the charmonium region have improved the cross section between 
3.7 and 4.8 GeV.
Since the widths of the poles are very small compared to the masses,
the narrow-width approximation provides an excellent description
of these states.
To model the contributions above the 6th resonance in the bottomium system
we use the assumption of quark-hadron-duality and integrate the
theoretical spectral density above $\sqrt{s_0}=11.0 \pm 0.2$ GeV. 
In the charmonium system
we include the two lowest resonances, the BES data and the theoretical
spectral density above 4.8 GeV.


\section{Numerical analysis}

The theoretical part of the correlator contains the poles of the 
Green's function, the spectral density above threshold and the condensates.
For high velocities the spectral density is well described
by the perturbative expansion whereas the resummed spectral density 
gives a good description for low values of $v$.
Therefore we construct a theoretical spectral density in the full 
energy range which includes the essential information in both
regions of $v$. For a more detailed discussion the reader is referred
to \cite{e:02}. Now we discuss the most important points in the
numerical analysis of the bottomium and charmonium systems respectively.


\subsection{Bottom mass}

First we discuss the values of $\xi$ and $n$. Since the bottom quark is 
relatively heavy, even for
$\xi=0$ the nonrelativistic and perturbative expansions converge
reasonably well. Nevertheless, the contributions from the poles of the
Green's function still dominate the theoretical part. To reduce their
influence and to spread the theoretical contributions more equally
among the poles, the resummed spectral density and the perturbative
spectral density we must choose a higher $\xi$. However, for 
$\xi>1$ the moments loose sensitivity on the mass and
the error from the input  parameters increases. Therefore we 
use a central value of
$\xi=0.5$ and vary $\xi$ between $0\leq \xi\leq 1$. 
Since the relevant scale for the evaluation point is the lowest
bound state energy, values of $\xi=0,0.5$ or 1 already correspond
to well separated evaluation points. 
For $n$ we use a range of  $5\le n\le 10$ where the theoretical
expansion and the  phenomenological uncertainty are under control.

As central values for our scales we have selected
$\mu_{soft}= 2.5$ GeV, $\mu_{fac}= 3.5$ GeV and 
$\mu_{hard}= 5.0$ GeV. For the error estimate we vary these
values between 
$2.0 \ \mbox{GeV} \leq \mu_{soft}\leq 3.5\ \mbox{GeV}$,
$2.0 \ \mbox{GeV} \leq \mu_{fac\ }\leq 5.0\ \mbox{GeV}$ and
$2.5 \ \mbox{GeV} \leq \mu_{hard}\leq 10.0\ \mbox{GeV}$.
For the separation scale $\mu_{sep}$ which appears as additional parameter
in the definition of the PS-mass we employ a value of
$\mu_{sep}=2.0\pm 1.0$ GeV.

\begin{figure}
\begin{center}
\vspace*{-7mm}
\includegraphics[height=\figwidth,width=\figheight,angle=-90]{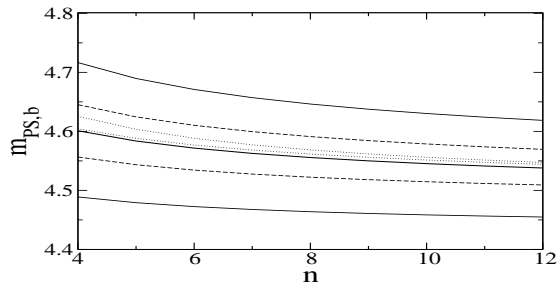}
\vspace*{-10mm}
\caption{\label{fig:8.b}
Thick solid line: central PS-mass;
thin solid lines: $m_{PS,b}$ for  $\mu_{soft}=2.0$
and 3.5 GeV;
dashed lines: $m_{PS,b}$ for  $\mu_{fac}=2.0$
and 5.0 GeV;
dotted lines: $m_{PS,b}$ for  $\mu_{hard}=2.5$
and 10.0 GeV.}
\vspace*{-10mm}
\end{center}
\end{figure}
The analysis is performed independently in the pole- and PS-scheme.
In figure \ref{fig:8.b} we have plotted the PS-mass as a function of $n$
and the influence of the scale variations. The largest contribution
to the error comes from the soft scale. 
Adding the errors from all input parameters quadratically, our final 
results for the masses are
\begin{eqnarray}
  \label{eq:8.e}
  M_b&=&4.984 \pm 0.125 \ \mbox{GeV}\,,\nn\\
  m_{PS,b}(2.0\ \mbox{GeV})&=& 4.561 \pm 0.112\ \mbox{GeV}\,,\nn\\
  m_{b}(m_{b}) &=& 4.241 \pm 0.098 \ \mbox{GeV}\,.
\end{eqnarray}


\subsection{Charm mass}

As in the bottom case we use $\xi=0.5$. At this value the pole
contributions still represent the dominant part. In principle one
would like to choose a higher value where the theoretical expansions
converge better. 
However, the contribution from the theoretical poles varies significantly
with the scales; for $\xi\gsim 1$ the mass depends too strongly on these
variations. Thus we again use a range of $0\leq \xi\leq 1$.
Since the perturbative expansions converge more slowly than for the upsilon
we restrict the analysis to smaller values of $4\leq n\leq 7$.
As central values for our scales we have selected 
$\mu_{fac}= 1.45$ GeV, $\mu_{hard}= 1.75$ GeV and values of 
$\mu_{soft}= 1.2$ GeV and $\mu_{soft}= 1.1$ GeV in the 
pole- and PS-schemes respectively.
For the error estimate we have varied the scales between
$1.4\ \gev\leq \mu_{hard}\leq 2.5 \ \gev$, 
$1.2\ \gev\leq \mu_{fac}\leq 1.65 \ \gev$ and
$1.1(1.0)\ \gev\leq \mu_{soft}\leq 1.35(1.25) \ \gev$.
For the separation scale we choose $\mu_{sep}=1.0\pm 0.2\ \gev$. 
\begin{figure}
\begin{center}
\vspace*{-7mm}
\includegraphics[height=\figwidth,width=\figheight,angle=-90]{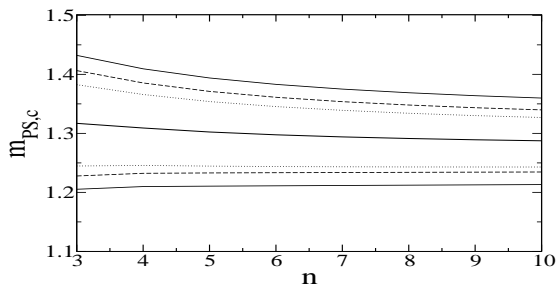}
\vspace*{-10mm}
\caption{\label{fig:9.b}
Thick solid line: central PS-mass;
thin solid lines: $m_{PS,c}$ for  $\mu_{soft}=1.0$
and 1.25 GeV;
dashed lines: $m_{PS,c}$ for  $\mu_{fac}=1.2$
and 1.65 GeV;
dotted lines: $m_{PS,c}$ for  $\mu_{hard}=1.4$
and 2.5 GeV.}
\vspace*{-10mm}
\end{center}
\end{figure}
In figure \ref{fig:9.b} we have plotted the PS-mass and the
corresponding error from the scales.
Finally we obtain the masses:
\begin{eqnarray}
  \label{eq:9.e}
  M_c&=&1.754 \pm 0.147 \ \mbox{GeV}\,,\nn\\
  m_{PS,c}(1.0\ \mbox{GeV}) &=& 1.300 \pm 0.124\ \mbox{GeV}\,,\nn\\
  m_{c}(m_{c}) &=& 1.188 \pm 0.106 \ \mbox{GeV}\,.
\end{eqnarray}


\section{Conclusions}

The method of QCD sum rules is a very powerful
tool to extract the masses since - by the choice of $n$ and $\xi$ -
it can react very sensitive to threshold. Thus, large theoretical
uncertainties only lead to a relatively small shift in the masses.
We have tried to develop a consistent framework to describe the physics
of the relevant energy region. 
The masses show a
stable behaviour over a large range of $n$ and the 
dominant uncertainties originate from the threshold expansion of NRQCD.

\vspace*{-2mm}
                 
\bigskip \noindent
{\bf Acknowledgements}

\noindent
I would like to thank S. Narison for the invitation to this
pleasant and interesting conference.
This work has been supported in part by TMR, EC
contract No. ERB FMRX-CT98-0169, by MCYT (Spain) under grant
FPA2001-3031, and by ERDF funds from the European Commission.
I thank the Deutsche Forschungsgemeinschaft for financial support.


\end{document}